# Controlled chemical vapor deposition for synthesis of emerging Mo(W)$Te_2$ systems

Ya Deng [a,†], Zi-Yi Han [b,c,†], Yao Wu [d,†], Kongyang Yi [a], Ya-Ning Ren [b,c], Dundong Yuan[e], Chao Zhu [e,*], Lin He [b,c,*] , Zheng Liu [a,f,*]

[a] *School of Materials Science and Engineering, Nanyang Technological University, Singapore 639798, Singapore*
[b] *Center for Advanced Quantum Studies, School of Physics and Astronomy, Institute for Advanced Study, Beijing Normal University, Beijing 100875, China*
[c] *China and Key Laboratory of Multiscale Spin Physics, Ministry of Education, Beijing 100875, China*
[d] *National University of Singapore; Singapore 117544, Singapore*
[e] *SEU-FEI Nano-Pico Center, Key Laboratory of MEMS of Ministry of Education, School of Integrated Circuits, Southeast University, Nanjing 210096, China*
[f] *IRL 3288 CINTRA (CNRS-NTU-THALES), Nanyang Technological University, 637553, Singapore*

---

**Abstract**

The Group-VI transition metal ditellurides offer a rich platform for correlated and topological phenomena, yet their structural polymorphism and instability complicate the creation of single crystals and heterointerfaces. Here, we introduce a confined-space chemical vapor deposition (CVD) strategy that lowers the growth temperature window and, when combined with tailored precursor configurations and stepwise thermal ramps, enables the deterministic synthesis of high-quality single crystals, alloys, and lateral/vertical heterostructures. High-resolution aberration-corrected STEM provides atomic characterization of lattice-matched Mo(W)$Te_2$ lateral heterostructure, revealing nearly atomically sharp, compositionally well-defined seamless boundaries. This approach avoids the thickness nonuniformity and structural limitations commonly associated with exfoliated samples, enabling reproducible fabrication of clean heterointerfaces and establishing a nearly ideal *in-situ* experimental system. Furthermore, scanning tunneling microscopy and spectroscopy (STM and STS) enable direct imaging of the seamless boundaries in Mo(W)$Te_2$ lateral heterostructures, while uncovering their distinct real-space distributions of the local density of states. Our results establish a scalable pathway for engineering crystalline Te-based structures with controlled geometry and stacking, providing an essential step toward quantum and topological device platforms based on the transition metal ditellurides family.

*Keywords: 2D materials, CVD growth, transition metal ditellurides, confined-space, heterostructures*

## Introduction

Two-dimensional (2D) transition metal dichalcogenides have emerged as a versatile platform for investigating novel quantum phenomena and enabling next-generation electronic and spintronic devices [1–3]. Within this class, the Group VI transition metal ditellurides family has attracted significant attention due to their complex phase structures, diverse electronic and topological properties [4–7]. Notably, MoTe2 and WTe2 serve as representative systems for exploring a wide range of emerging quantum states, including type-II Weyl semimetals, superconductors, topological insulators, and strongly correlated phases [8–17]. Moreover, MoTe2/WTe2 lateral heterostructures have been reported as a promising platform for integrating Majorana qubits, offering potential for future topological quantum computation [18]. Meanwhile, MoTe2 and WTe2 are also of great interest also due to their shared similar distorted octahedral structure but exhibit distinct topological electrical characteristics. Constructing heterostructures from these two phases enables side-by-side comparison of structurally analogous materials, also provides a nearly ideal platform to investigate interfacial coupling between distinct topological states. However, traditional top-down synthesis methods inherently limit the fabrication of heterostructures. Furthermore, in Mo(W)Te2 systems, the structural instabilities introduced during top-down approaches such as mechanical exfoliation can influence the layered stacking order, thereby limiting the investigation of their intrinsic properties [19].

To overcome these limitations, direct bottom-up chemical vapor deposition (CVD) offers a promising alternative for synthesizing high quality MTe2 (where M = Mo or W) single crystals and complex heterostructure layers, eliminating the need for mechanical exfoliation. However, the controlled synthesis of Mo(W)Te2 systems with well-defined structures remains challenging due to its thermodynamic instability and complex phase behavior. In particular, the structural similarity between 1T′-MoTe2 and

1Td-$WTe_2$, including their nearly identical lattice constants, promotes alloy formation but poses challenges for the direct growth of clean, compositionally sharp lateral or vertical heterostructures, thereby hindering the fabrication of abrupt heterointerfaces required for advanced device applications [20]. Therefore, establishing a versatile growth strategy capable of precisely controlling the growth of Mo(W)$Te_2$ family including single crystals, alloys, and heterostructures, provides a nearly ideal *in-situ* experimental platform for precise side-by-side study of transition metal ditellurides systems and for expanding the landscape of unprecedented material combinations with emergent properties.

Here, we present a robust and highly controllable confined growth strategy that enables the synthesis of Mo(W)$Te_2$ single crystals at lower growth temperatures, while also allowing for the fabrication of heterostructures and alloys with excellent spatial and compositional precision. Employing high-resolution scanning transmission electron microscopy (STEM) and scanning tunneling microscopy/spectroscopy (STM/STS), we directly visualize the atomic and electronic structures across the resulting heterojunctions, thereby uncovering sharply defend, seamless boundaries and distinct real-space local density of states (LDOS) patterns. This work not only provides a broadly applicable approach for constructing nearly atomically sharp heteroepitaxial interfaces in lattice-matched 2D systems but also establishes a versatile platform for investigating emergent physical phenomena in transition metal ditellurides.

We synthesized high-quality Mo(W)$Te_2$ systems using a confined space chemical vapor deposition (CVD) strategy integrated with a glove box system, as illustrated in Fig. S1. This setup provides a fully inert environment from precursor loading to post-growth characterization, minimizing oxidation and contamination. By employing spatially separated precursors (Fig. 1a), precisely controlling the growth temperature and atmospheric conditions, and temporally decoupling the nucleation and growth stages of each material, we achieved highly controllable synthesis across Mo(W)$Te_2$ systems. The morphology and structures of synthesized Mo(W)$Te_2$ members are schematically illustrated in Fig. 1b–e. The synthesis of pure-phase single crystals serves as the foundational step and is relatively straightforward, as demonstrated in our previous work [7]. In general, the growth of $MoTe_2$ and $WTe_2$ is primarily governed by growth temperature and the carrier gas atmosphere. Previous studies reported the growth temperatures of 780 ◦C for growing $MoTe_2$ and 820 ◦C for growing $WTe_2$ single crystals [7]. In this work, we employ a confined growth mode as shown in Fig. S2 and optimized precursor configuration, which significantly reduced both the growth temperature and duration. For example, the growth conditions for $MoTe_2$ were lowered from 780 ◦C for 5 min to 680 ◦C for 2 min. Optical images confirming the successful synthesis of $MoTe_2$ and $WTe_2$ using this optimized strategy are presented in Fig. S3. This synthesis strategy effectively decouples the growth stages of the two materials, providing a foundation for the fabrication of more complex structures. However, confinement alone is not sufficient to achieve heterostructures. Under previous one-pot growth conditions, the nucleation processes of $MoTe_2$ and $WTe_2$ interfere with each other, leading to the formation of $Mo_xW_{1-x}Te_2$ alloys with varying doping levels, even when precise temperature control is applied. The calculations shown in Note S1 further reveal that the lattice parameters of $MoTe_2$ and $WTe_2$ are well matched, making it easier to obtain $Mo_xW_{1-x}Te_2$ alloys in the final product.

To achieve heterostructure formation, we developed a one-pot, twostep chemical vapor deposition (CVD) strategy and the detailed synthetic strategy is illustrated in Fig. 2. First, we maintain limited separation between the $MoO_3$ and $WO_3$ precursors and utilize spatial confinement within the corundum boat to help isolate the reaction regions of the two materials. Notably, the distances between precursors (denoted as $l$) and their vertical distances from the substrate surface (denoted as $h$) play a crucial role in determining the resulting growth products. A small $l$ (<6 mm) leads to excessive intermixing and alloy formation, while a large $l$ (>15 mm) results in the formation of isolated single crystals rather than lateral heterostructures. Furthermore, the parameter $h$ controls the availability of carrier gas near the substrate surface, thereby influencing whether the reaction can be effectively sustained. In addition, we developed a two-step growth strategy: At 680 ◦C, $MoTe_2$ was selectively growth first, while $WTe_2$ had not yet reached the required reaction temperature for nucleation (Fig. 2a). After holding this temperature to fully consume the $MoO_3$ precursor, the growth process was ramped to a higher temperature and optimized the gas atmosphere for $WTe_2$ growth. The optimized temperature staging, combined with the independent controlled carrier gas flow (Fig. 2b), enables epitaxial growth of heterostructures. Notably, this growth sequence differs from the conventional strategy commonly used for lateral heterostructures, in which the material requiring a higher growth temperature is typically grown first. In this work, the $MoTe_2$ was grown first at a lower temperature because the closely matched lattice constants and overlapping growth temperature windows of $MoTe_2$ and $WTe_2$ promote $Mo_xW_{1-x}Te_2$ alloy formation when both metal precursor vapors are simultaneously present at higher temperatures. In addition, the optimized reaction temperature for STEP 2 is different for growing in-pane and vertical heterostructures. Lower growth temperatures allow lattice arrangement to be maintained at the interface during subsequent nucleation, thus enabling lateral stitching without disrupting crystallographic coherence, while higher temperature helps overcome the energy barrier at the van der Waals interface, promoting vertical heterostructure formation. Fig. 2c displays representative samples synthesized under different stages, ranging from individual single crystals to $MoTe_2$/ $WTe_2$ lateral heterostructures. Compared to the noticeable colour variations observed in $MoTe_2$ with different thicknesses in the left panel of Fig. 2c, the colour contrast at the heterostructure boundary appears relatively weak. This suggests that the lateral heterostructure likely exhibits uniform thickness across the junction, with the contrast arising from differences in material composition on either side. By employing precursor configuration, spatial confinement, and carefully designed synthesis parameters, we achieved controlled and orderly growth, enabling the successful synthesis of complex heterostructures within the $MTe_2$ system. It demonstrates a viable pathway to stabilize lateral heterojunctions between highly miscible systems, providing a novel strategy for the controlled preparation of heterostructures from otherwise alloy-prone material pairs. More experimental details and optical images of other heterostructures are provided in the "Materials and Methods" section and Figs. S4–S8a.

Next, we mainly focus on lateral heterostructures of 1T′-$MoTe_2$ and 1Td-$WTe_2$, as these can only be obtained through a direct growth method, and to our knowledge, are reported here for the first time. As shown in the optical image in Fig. 3a and schematic diagram in Fig. 3b, the central region of the sample flake exhibits a dark purple, whereas the edges appear lighter in colour. To determine whether this colour contrast originates from compositional differences, we performed Raman spectroscopy. Fig. 3c presents Raman spectra collected from three representative points (P1 to P3), spanning from the center to the edge. The central region (P1) shows a characteristic Ag peak of $MoTe_2$ at 260 $cm^{-1}$, while the edge region (P3) exhibits the A1 peak of $WTe_2$ at 210 $cm^{-1}$. The intermediate region displays features from both materials, indicating the presence of the heterostructure. Furthermore, the atomic force microscopy (AFM) image in Fig. 3d reveals uniform thickness across both sides of the interface, confirming that the heterostructure is lateral rather than vertically stacked. Additional Raman mapping in Fig. 3e–g further confirms the uniform distribution of the two materials, revealing a well-defined, compositionally sharp, and structurally clean interface throughout the lateral heterostructure. More Raman characterization results of the other $MTe_2$ structures are presented in Figs. S4–S8, confirming the Mo(W)$Te_2$ single crystals, vertical heterostructures, and alloy systems.

The atomic structure and composition of the lateral heterostructures were further investigated by scanning transmission electron microscopy (STEM) and energy-dispersive X-ray spectroscopy (EDX) characterization. The EDX maps of Mo, W, and Te shown in Fig. 4a and Fig. S9 clearly confirm the distribution of each constituent element within the $MoTe_2$/ $WTe_2$ lateral heterostructure and $MTe_2$ other structures. As shown in Fig. 4b, the annular dark-field (ADF)-STEM image overlaid with the corresponding false-coloured atomic number (*Z*)-contrast maps highlights the interface between $WTe_2$ and $MoTe_2$, revealing a seamless junction. The inset shows the intensity line profile extracted across the interface, demonstrating a nearly atomically sharp transition occurring in ~2 nm region. The high-resolution ADF-STEM images in Figs. 4c and S10, S11 further provide direct visualization of the crystal structures on both sides of the heterointerface, with the insets resolving the atomic structures of monolayer $WTe_2$ and $MoTe_2$. Owing to the *Z*-contrast feature of ADFSTEM imaging, the heavier tungsten atoms appear distinctly brighter, whereas the lighter molybdenum atoms on the right side of the heterostructure exhibit comparatively lower intensity, enabling direct visualization of the compositional distribution at the atomic scale. These results clearly demonstrate the formation of a laterally seamless and nearly atomically sharp heteroepitaxial interface. First-principles calculations have predicted subtle differences between $MoTe_2$ and $WTe_2$ in terms of their electronic structure and related phenomena [20,21]. However, experimentally verifying these distinctions remains highly challenging, as conventional STM measurements can be influenced by extrinsic factors such as sample thickness, substrate effects, and surface quality. The $MoTe_2$/$WTe_2$ lateral heterostructure reported in this work provides a nearly ideal *in-situ* controlled experimental platform for performing a precise, side-by-side study of $MTe_2$ systems. The atomically sharp interface enables direct side-by-side comparison of electronic structures. Beyond probing the intrinsic electronic properties of individual $MTe_2$ materials, particular interest lies in the interfacial states that emerge within this lattice-matched lateral heterostructure. To probe these effects, we performed high-resolution STM and STS measurements on monolayer sample supported on Au (111) substrate (see sample structure in Fig. S12).

Large-scale topographs (Figs. 5a and S13) directly reveal a long and straight interface in the $MoTe_2$/$WTe_2$ lateral heterostructure. The boundaries may possess certain chemical activity, as clusters or molecules have been observed to adsorb randomly onto them in our STM measurements. The abrupt topographic transition across the interface indicates an abrupt chemical transition, which is consistent with observations in previously reported heterostructures on sapphire and graphite substrates [22,23]. A zoomed-in STM image of a straight segment (Fig. 5b) reveals seamless stitching of the two lattices. Meanwhile, atomic-resolution images of $MoTe_2$ and $WTe_2$ on either side of the junction (Fig. 5c), along with their corresponding fast Fourier transforms (FFT, insets in Fig. 5c), confirm the 1T' lattice with one dimensional (1D) metallic atomic chains aligned parallel to the interface, in good agreement with STEM observations (Fig. 4c). To explore the electronic structure, we performed spatially resolved STS measurements across the heterojunction. Fig. 5d shows the site-dependent STS spectra collected along an 80 nm line crossing the interface. Under identical experimental conditions, the STS spectra on either side of the junction are analogous yet exhibit subtle differences, which agrees well with the calculated DOS of monolayer $MoTe_2$ and $WTe_2$ reported in previous studies [24,25], suggesting that the Au (111) substrate does not obscure the intrinsic electronic states. It also indicates that our heterostructure serves as a pristine experimental benchmark, offering unambiguous experimental data for validating the theoretical calculations. Notably, the interfacial STS spectra exhibit a dip-like feature near the $WTe_2$ side, which can be attributed to band bending driven by interfacial charge redistribution. This quasi-1D band modulation may further give rise to partially confined interfacial electronic states. To directly visualize the electronic contrast across the interface, we mapped the LDOS in real space across the junction (Fig. 5e). The results reveal differences in electronic intensity between the $MoTe_2$ and $WTe_2$ regions at the bias voltages corresponding to the STS peaks in Fig. 5d, with a clear boundary resolved between the two regions (see additional d$I$/d$V$ maps in Fig. S14). This distinct electronic character provides new opportunities for exploring novel quantum states at micron-scale, atomically sharp 1D boundaries within 2D systems.

In summary, we have developed a controlled synthesis strategy for Mo(W)Te2 systems, enabling the precise fabrication of high-quality single crystals, heterostructures, and alloys. Moreover, we report the successful synthesis of the lateral MoTe2/WTe2 heterostructure, enabling a nearly atomically sharp and seamless heteroepitaxial interface between two distinct but lattice-matched quantum materials within a single atomic layer. Using high-resolution STEM and STM, we directly visualized local electronic and structural contrasts across the interface, confirming the theoretical calculation regarding the subtle yet critical differences between MoTe2 and WTe2. This work also provides valuable insights into novel synthesis strategies for 2D heterostructures and paves the way for exploring emergent physical phenomena in the promising transition metal ditellurides family

## Methods

*Synthesis of atomically thin Mo(W)Te2 systems*

All synthesis reactions were carried out in a 1-inch quartz tube furnace. The metal oxide precursors were placed in a custom corundum boat with a depth of 3 mm (Fig. S2), positioned at the center of the tube. The 280 nm SiO2/Si substrate was positioned face down on the corundum boat directly opposite the precursor, as illustrated in Fig. S2. After synthesis, the samples were collected directly from the integrated glovebox port (Fig. S1). This setup ensured the synthesized air-sensitive Mo(W)Te2 systems remained under a continuous inert atmosphere for all subsequent characterization or transfer processing. Detailed synthesis parameters varied depending on the material.

*Synthesis of MoTe2 single crystal*: To synthesize MoTe2 single crystal, 10 mg of MoO3 powder mixed with 1 mg of NaCl was used and placed in the corundum boat as the precursor. The carrier gas flow was set to H2/ Ar (2/100 sccm) during the reaction stage. The furnace was ramped to 650 ◦C in 15 min, maintained at this temperature for 2 min, and then naturally cooled to room temperature.

*Synthesis of WTe2 single crystal*: To synthesize WTe2 single crystal, 10 mg of WO3 powder mixed with 1 mg of NaCl served as the precursor. The carrier gas flow was set to H2/Ar (5/100 sccm) in the reaction stage. The furnace was ramped to 750 ◦C in 18 min, maintained at this temperature for 2 min, and naturally cooled to room temperature.

*Synthesis of MoxW1-xTe2 alloys*: To synthesize Mo*x*W1-*x*Te2 alloys, 10 mg each of MoO3 and WO3 powders were mixed with 1 mg of NaCl and used as the precursor. The carrier gas flow was set to H2/Ar (5/100 sccm) in the reaction stage. The furnace was ramped to 750 ◦C in 18 min, maintained at this temperature for 2 min, and naturally cooled to room temperature.

*Synthesis of MoTe2/WTe2 heterostructures*: The synthesis of MoTe2/ WTe2 heterostructures required additional control. The carrier gas flow was optimized independently for the two growth steps (detailed in Fig. 2b). In STEP 1, a relatively low H2 flow was used for MoTe2 growth while avoiding excessive uncontrolled nucleation. In STEP 2, the H2 flow 6 was increased to promote WTe2 growth, which requires a stronger reducing environment and a higher temperature. Besides the precise control of temperature and carrier gas flow (detailed in Fig. 2b), the ratio and placement of precursors were also critical. To prevent excessive intermediate reactions and uncontrolled growth, the amount of salt was halved in both precursor mixtures. Specifcally, 10 mg of MoO3 with 0.5 mg of NaCl was used for STEP 1, and 10 mg of WO3 with 0.5 mg of NaCl for STEP 2. The spatial separation between the two precursors, denoted as $l$ in Fig. 2a, was optimized to 10 mm, allowing for synergistic yet noninterfering vapor transport. For the growth of lateral heterostructure, the temperature was gradually ramped to 740 ◦C, the optimized growth condition for STEP 2, to facilitate the epitaxial growth of WTe2 along the existing MoTe2 edge. Conversely, for the growth of vertical heterostructure, the temperature was rapidly increased to 780 ◦C in STEP 2 to overcome the energetic barriers for out-of-plane stacking. After the reaction of STEP 2, the hydrogen flow was shut off, and the furnace was rapidly cooled down to room temperature to preserve the structure.

*AFM and Raman characterization* Atomic force microscopy (AFM) characterization was conducted using an Asylum Research Cypher AFM operating in tapping mode. Raman measurements were performed using a WITec Alpha 200R confocal Raman system with a 532 nm excitation laser, and the laser power was kept below 1 mW to prevent overheating of the samples. For Raman mapping measurements, the samples were coated with PMMA to avoid oxidation during data acquisition.

*Electron microscopy characterization* The TEM/STEM samples were prepared using a direct wet-etching method. A Quantifoil Au TEM grid was placed face down onto the target region of the sample and held in position with tweezers. A drop of 3 mol/L KOH solution was then applied to the target area and left for 30 s to etch the SiO2 layer. The grid was subsequently rinsed in deionized water to remove residual KOH etchant. Finally, the samples were treated with a plasma cleaner in an Ar/H2 atmosphere (10 W, 2 min) to remove hydrocarbon contamination, completing the preparation of the grid samples for characterization. STEM characterization was performed using the JEOL ARM200F microscope equipped with a CEOS aberration corrector and operated at 80 kV. The TEM characterization was conducted using a JEOL 2200FS at 200 kV. The SEM characterization was conducted using a JEOL JSM-7600F at 15 kV.

*STM characterization* The STM/STS measurements were performed in low-temperature (77 K) and ultrahigh-vacuum (~10- 10 Torr) scanning probe microscopes (USM-1400) from UNISOKU. The STM tips were prepared by chemical etching tungsten wire. The differential conductance (d*I*/d*V*) measurements were taken by a standard lock-in technique with an ac bias modulation of 5 mV and 793 Hz signal applied to the tunneling bias.

*STM sample preparation* The STM samples were prepared using a direct wet-etching method similar to STEM samples. The Au (111) substrate was placed face down onto the target region of the sample and held in position with tweezers. A drop of 3 mol/L KOH solution was then applied to the target area and left for 30 s to etch the SiO2 layer. The substrate was subsequently rinsed with deionized water to remove residual KOH etchant

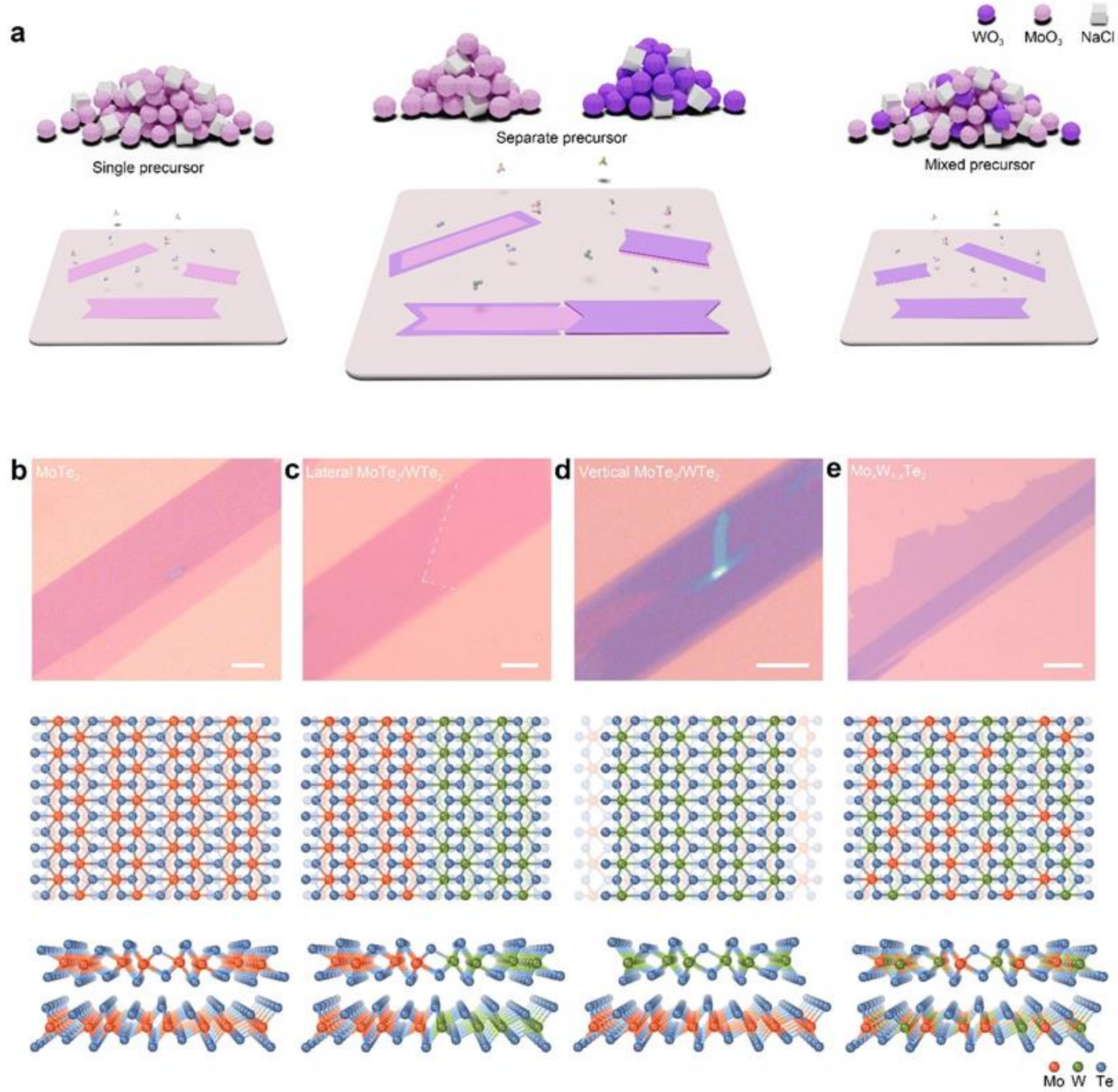

Fig. 1. Synthesis and morphologies of diverse $Mo(W)Te_2$ systems. (a) Schematic illustration of three types of precursor configurations for synthesizing $Mo(W)Te_2$ systems. (b-e) Typical optical images and corresponding atomic structural models (viewed from top and front directions) of synthesized single crystal flakes (b), lateral heterostructures (c), vertical heterostructures (d), and alloy systems (e). Red, green, and blue balls represent Mo, W, and Te atoms respectively. Scale bars: 10 µm.

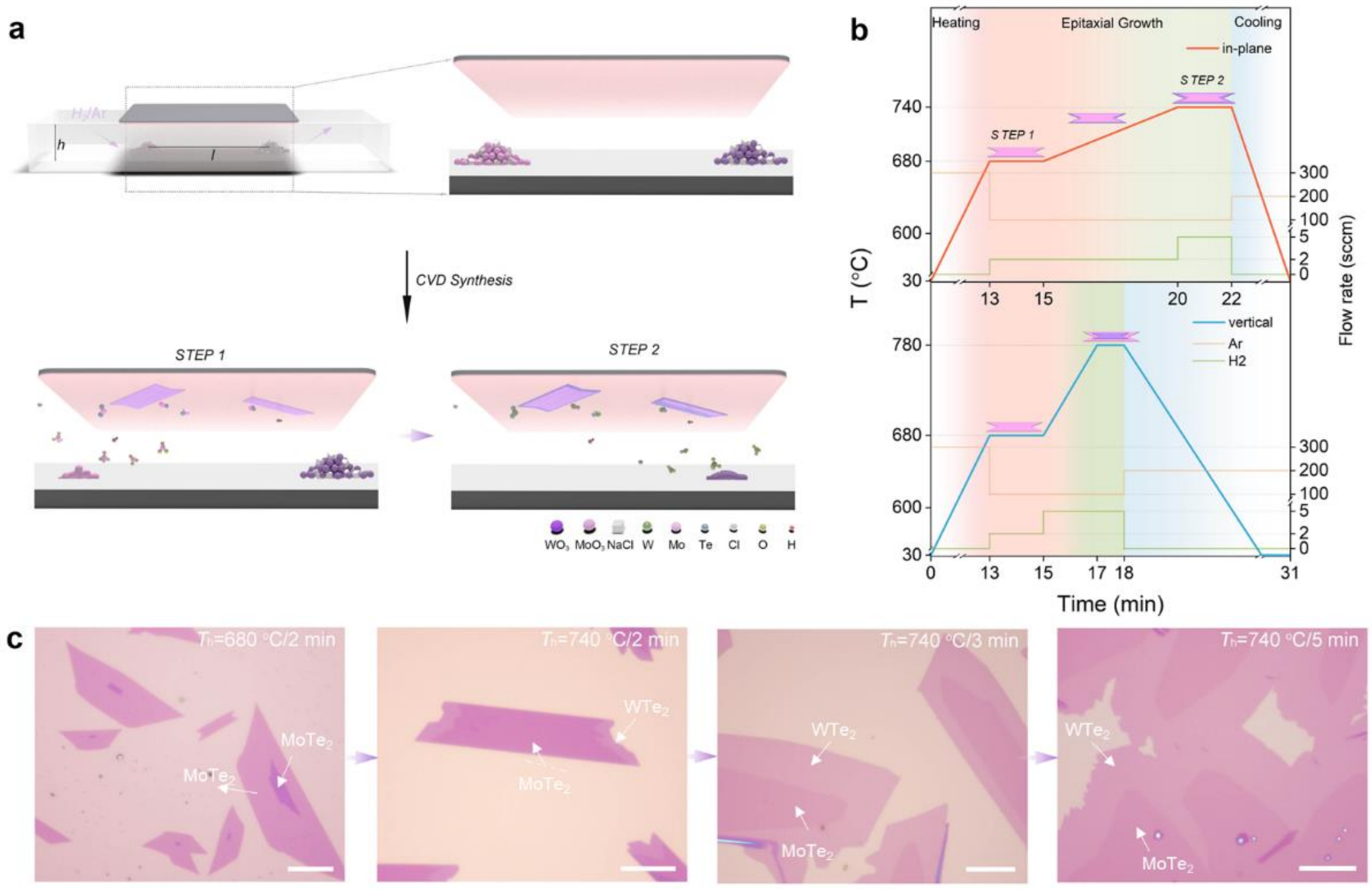

Fig. 2. Growth mechanism of Mo(W)$Te_2$ heterostructures. (a) Schematic illustration of the confined-space chemical vapor deposition setup employing separated precursors. The *l* indicates the separation between the precursors, while the *h* represents the vertical spacing between the precursor and the substrate. (b) Growth parameters used for synthesizing $MoTe_2$/$WTe_2$ lateral and vertical heterostructures. (c) Optical microscopy images of samples collected after stopping the synthesis process at STEP1 and STEP2 with reaction times of 2, 3, and 5 minutes, revealing the morphological evolution from isolated $MoTe_2$ single crystals to $MoTe_2$/$WTe_2$ lateral heterostructures. Scale bars: 10 µm.

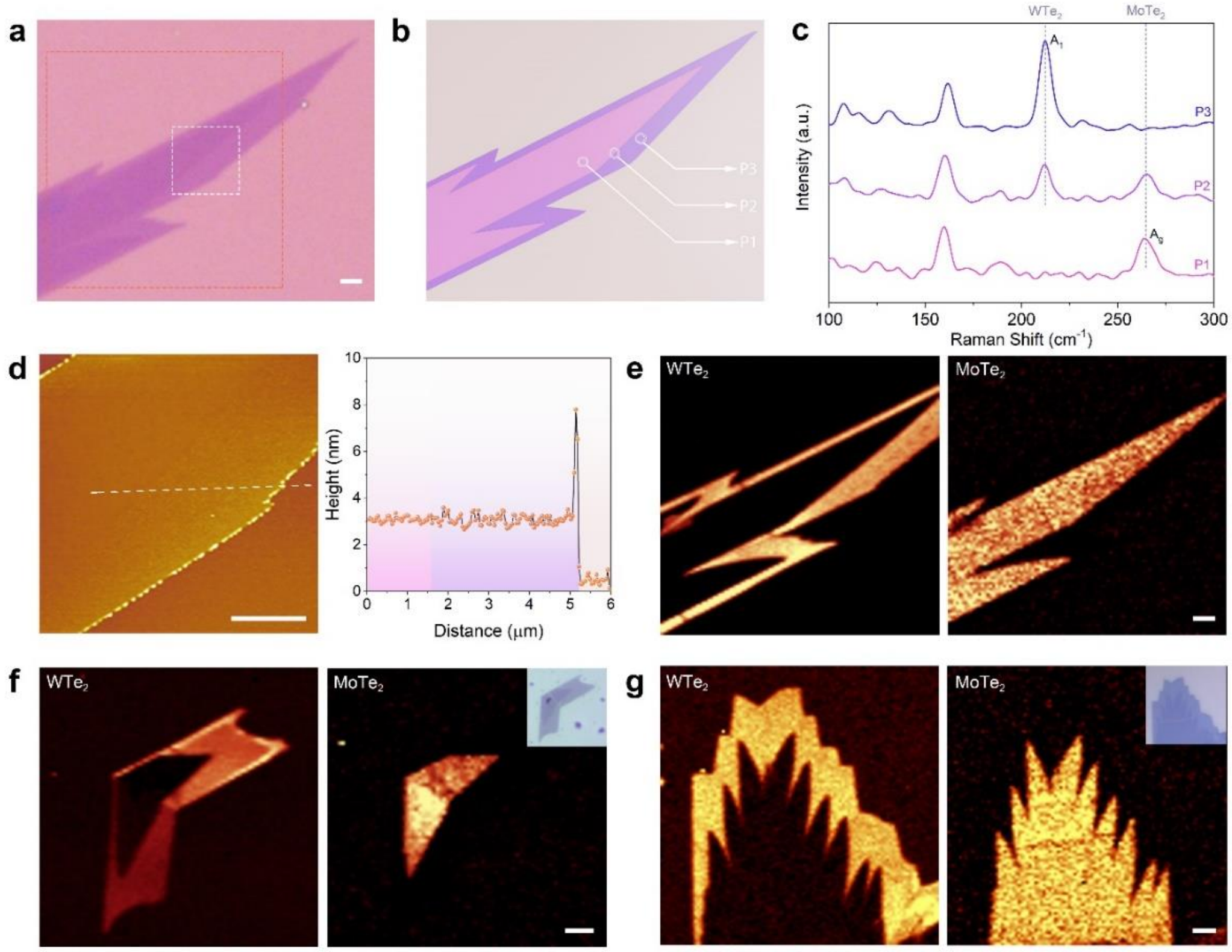

Fig. 3. Morphologies and Raman characterization of lateral $MoTe_2$/$WTe_2$ heterostructures. (a-b) Optical image (a) and schematic illustration (b) of a representative $MoTe_2$/$WTe_2$ heterostructure. (c) Raman spectra collected from positions P1 to P3, confirming the characteristic peaks of $MoTe_2$ (P1), $WTe_2$ (P3), and the interface (P2) regions. (d) Atomic force microscopy (AFM) image of the white dashed rectangle region in a. The right panel is the height profile along the white dashed line, highlighting the consistent height across the interface of the lateral heterostructure. (e) Raman $A_1$ and $A_9$ intensity mapping of the orange dashed rectangle region in a, revealing the $MoTe_2$ in the inner region and the $WTe_2$ at the edge region. (f-g) Raman mapping of PMMA-covered lateral heterostructures with corresponding optical images (inset). Scale bars: 2 μm.

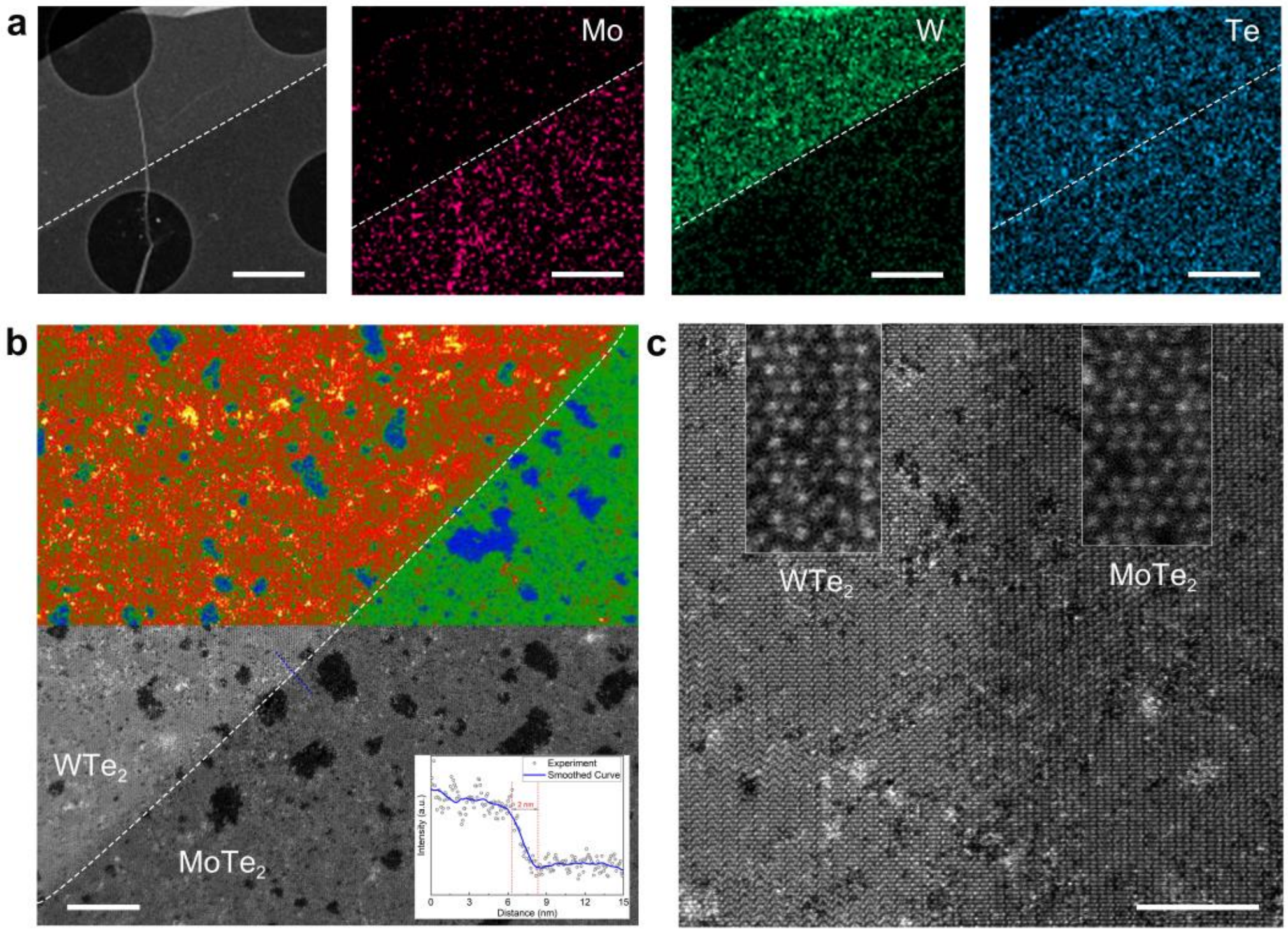

Fig. 4. Atomic structural and composition characterization of lateral $MoTe_2/WTe_2$ heterostructure. (a) Transmission electron microscopy (TEM) image of the heterostructure (left) and corresponding energy-dispersive X-ray spectroscopy (EDX) elemental maps for Mo (pink), W (green), and Te (blue), confirming the distribution of each element. (b) Overlay of ADF-STEM image (bottom) and corresponding false-colored Z-contrast image (top) showing the interface between $WTe_2$ and $MoTe_2$ of the bilayer heterostructure. The inset shows a line intensity profile along the blue line, demonstrating an atomically sharp and seamless heteroepitaxial interface. (c) High-resolution STEM image of the heterostructure interface. Insets revealing atomic-resolution images of the 1 nm wide bottom monolayer $WTe_2$ and $MoTe_2$ regions. Scale bar: 1 μm in (a), 20 nm in (b) and 5 nm in (c).

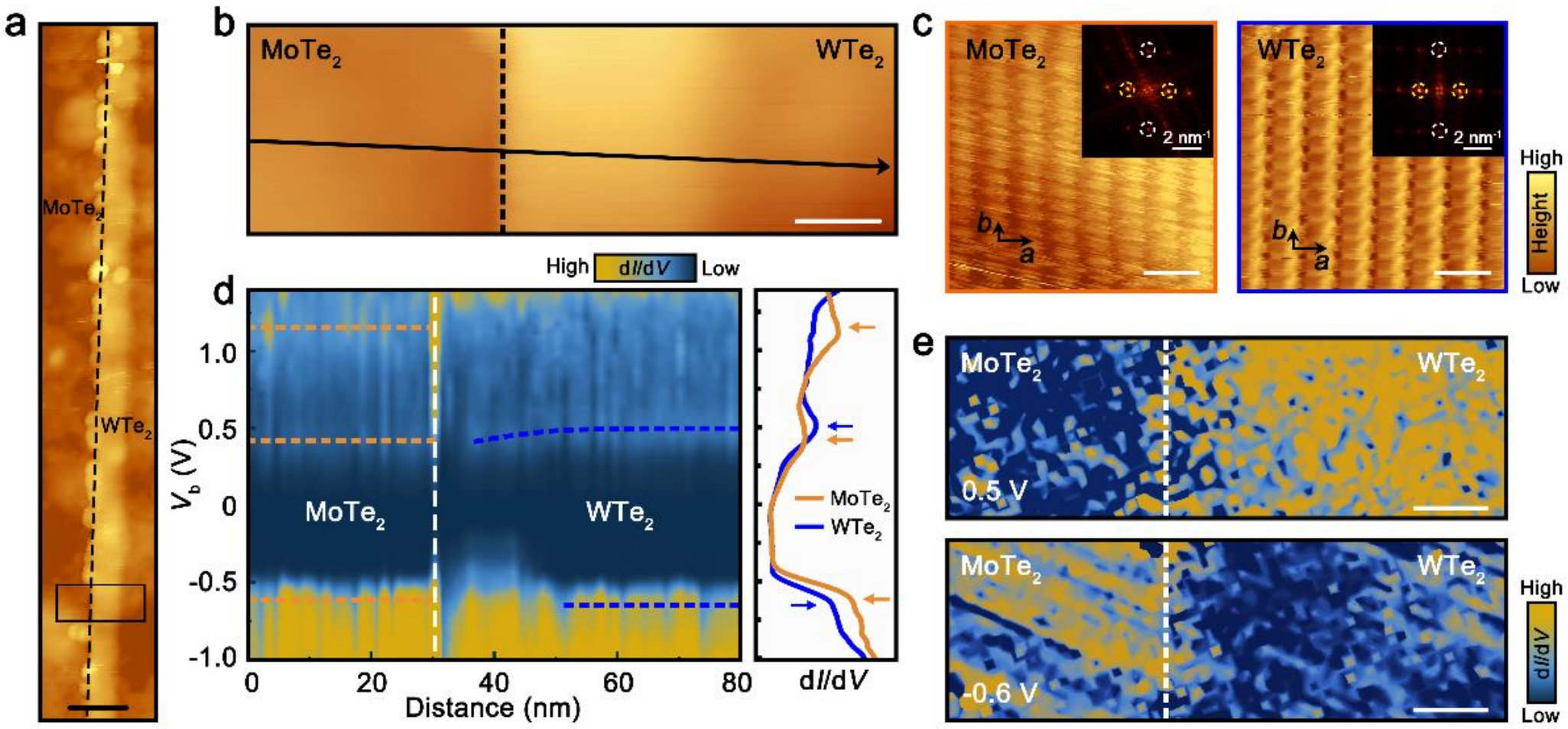

Fig. 5. STM characterization of a lateral $MoTe_2/WTe_2$ heterostructure. (a) Large-scale STM topography of a $MoTe_2/WTe_2$ heterostructure (bias voltage $V_b$ = 1.1 V, tunneling current I = 100 pA). The black dashed line marks the 1D interface of the heterojunction. (b) Zoomed-in view of the $MoTe_2/WTe_2$ heterostructure corresponding to the black box in panel a ($V_b$ = -1 V, I = 100 pA). (c) Atomic-resolution images of $MoTe_2$ (left panel, $V_b$ = 0.3 V, I = 100 pA) and $WTe_2$ (right panel, $V_b$ = 0.4 V, I = 100 pA) on the two sides of the heterostructure. Insets: corresponding FFT images. (d) Left panel: Spatially resolved STS spectra recorded along the black arrow in panel b (left panel). The interface of the heterostructure is marked by the white dashed line. The orange and blue dashed lines denote the $MoTe_2$ and $WTe_2$ density-of-states peaks, respectively. Right panel: Typical dI/dV spectra of $MoTe_2$ (orange line) and $WTe_2$ (blue line) regions. The arrows indicate their peak positions corresponding to the dashed lines in the left panel. (e) The dI/dV maps of the region in panel b acquired at $V_b$ = 0.5 V and -0.6 V (I = 200 pA). The white dashed lines mark the 1D interface of the heterostructure. Scale bars: 50 nm in (a), 10 nm in (b), 1nm in (c) and 10 nm in (e).